\documentclass[12pt, aps,amsmath,showpacs,superscriptaddress, pra]{revtex4-1}
\usepackage{dcolumn,bbm,amssymb,amsmath,ulem,indentfirst,amsthm,color,graphicx}
\begin{document}
\title{Finite-sample deviations and convergence in the statistics of Bohmian trajectory ensembles}

\author{Bingyu Cui}
\email{bycui@cuhk.edu.cn}
\affiliation{School of Science and Engineering, The Chinese University of Hong Kong (Shenzhen), Longgang, Shenzhen, Guangdong, 518172, P.R. China}
\author{Yanting Cao}
\affiliation{Department of Technology Management for Innovation, The University of Tokyo, 7-3-1 Hongo, Bunkyo-ku, Tokyo 113-8656, Japan}
\date{\today}

\begin{abstract}
\noindent We analyze finite-sample statistics of Bohmian trajectories for single spinless and spin-1/2 particles. Equivariance ensures agreement with $|\psi|^2$ in the quantum equilibrium limit, yet experiments and simulations necessarily use finite ensembles. We show that in regular flows (e.g., wavepackets or low-mode superpositions of eigenstates of harmonic oscillators) sample means and/or variances over modest $N$ are consistent with Born-rule moments. In contrast, degenerate superpositions of 3D oscillators with nodal barriers and chaotic Bohmian dynamics exhibit sensitive dependence on initial conditions and complex flow partitioning, which can yield noticeable finite-sample deviations in the mean and variance. For the spin-1/2 particle, both convective and Pauli currents conserve $|\psi|^2$, but they are associated with different velocity fields and thus might yield different finite-sample trajectory statistics. These findings calibrate the interpretation of trajectory-based uncertainty and provide practical guidance for numerical Bohmian simulations of spin and transport, without challenging the equivalence to orthodox quantum mechanics in the quantum equilibrium ensemble.
\end{abstract}

\pacs{}
\maketitle

\section{Introduction}
The exceptionally precise agreement with experiment achieved through quantum mechanics makes it the most successful and widely accepted theory for the microscopic world, where quantum particles behave in a completely different manner from their classical counterparts. In standard quantum mechanics, all information about a particle’s state is encoded in the wavefunction $\psi$, which evolves according to the Schr\"{o}dinger equation. According to the Born rule, the square amplitude of the wavefunction is interpreted as the probabilistic distribution. In orthodox quantum mechanics, physical quantities and the operators representing them, whose expected values are of physical interest, are often referred to as observables. It is meaningless to talk about the value of an observable unless a measurement, which causes the collapse of the wavefunction to an eigenstate of the corresponding operator, is performed \cite{vonNeumann2018}.

The randomness in observables is characterized by the so-called uncertainty principle \cite{BALLENTINE1970,Griffiths2018}, which sets a lower bound for certain pairs of observables obeying the commutation relation. The uncertainty principle rules out the simultaneous definition of positions and velocities at each instant. Following the Copenhagen interpretation, it has been comprehensively entrenched in many textbooks and lectures that the notation of a trajectory should be abandoned since it makes no sense to talk about the position or velocity of a particle at the same time. A thorough analysis of Heisenberg's argument \cite{heisenberg1983,heisenberg2013}, however, indicates that even Heisenberg himself never denied the existence of the trajectory based on this relation, and it is a matter of personal preference to accept or reject a trajectory-containing quantum theory \cite{Aristarhov2022}. The successful retrodiction of the particle's trajectory has been demonstrated in some experiments in past few decades \cite{Kocsis2011,Schleich2013}. 

In contrast, Bohmian mechanics posits that particles have well-defined positions at all times and evolve deterministically under velocity fields determined by $\psi$ \cite{Bohm1952,holland1993,Oriols2019}. Rather than being viewed as a numerical tool for obtaining predictions for laboratory experiments, the Bohmian interpretation of quantum mechanics provides a clear explanation of quantum randomness: the incapability of accessing the precise location of particles, whose motion is guided by the probability field $|\psi|^2$. Positions and momenta of quantum particles should thus be regarded as hidden variables \cite{Bohm1952}. Statistical predictions then arise from “quantum equilibrium hypothesis” (QEH): An ensemble of initial positions distributed by $|\psi|^2$ is transported by the guidance flow, maintaining equivariance so that the ensemble at later times again follows $|\psi|^2$ \cite{Durr1992,Oriols2019}. Mathematically, the QEH states that the distribution of particle positions in different experiments at time $t$ might be written as
\begin{equation}
    |\psi(\mathbf{x},t)|^2=\lim_{N\rightarrow\infty}\frac{1}{N}\sum_{n=1}^N\delta(\mathbf{x}-\mathbf{x}^n[t]),
\end{equation}
where $n=1,...,N$ are indices of the different trajectories $\mathbf{x}^n[t]$ belonging to different experiments.

In practice, however, only a finite amount of data is collected. It is then likely to record appreciable deviations between statistics based on a finite number of trajectories and the ideal results of standard quantum mechanics. This paper is not about challenging equivariance or Born-rule statistics. Rather, it addresses a pragmatic question highly relevant to numerical and experimental practice: How do finite ensembles of Bohmian trajectories—unavoidable in simulations and laboratory runs—approximate standard quantum statistics across different dynamical regimes, and when can finite-sample deviations be large enough to matter?

Two structural features of Bohmian guidance are central. First, nodes of the wave function $(\psi=0)$ create invariant barriers; trajectories cannot cross nodal surfaces, partitioning configuration space into nodal cells. Second, depending on the state, guidance can be stationary, quasi-periodic, or chaotic. In many systems (e.g., wavepackets or superpositions of a few eigenstates of harmonic oscillators) the flow is regular and mixing within a cell is gentle; finite samples tend to reproduce quantum moments robustly. However, in particular superpositions with degeneracies and nodal structures, the flow can exhibit sensitive dependence on initial conditions (positive Lyapunov exponents) and poor inter-cell mixing, potentially amplifying finite-sample deviations in low-order statistics for realistic sample sizes. Related work and positioning Bohmian chaos and nodal structures have been studied extensively in dynamical systems analyses of guidance flows, including for harmonic oscillators and other model systems \cite{Frisk1997,Konkel1998,Wu1999,Falsaperla2003,Valentini2005, Wisniacki2005, Wisniacki2006,Efthymiopoulos2007, Contopoulos2008,Efthymiopoulos2009,Borondo2009,Contopoulos2012,Cesa2016}. Our focus differs by quantifying the impact of these flow properties on finite-ensemble statistical estimates (means/variances) rather than on individual trajectory geometry or ergodicity. 

We also examine spin-1/2 dynamics, where the Pauli equation yields a continuity equation under multiple admissible currents (e.g., convective vs total/Pauli currents). Both preserve $|\psi|^2$ (equivariance) but define different velocity fields. This raises a question: To what extent can such choices affect finite-sample trajectory statistics for observables inferred via position readout? Prior discussions have examined convective vs Pauli current choices and their implications for trajectories \cite{holland1993,Colijn2002,Holland2003}; we complement this by comparing trajectory statistics using convective and Pauli currents, documenting differences in finite-sample behavior while preserving ensemble equivalence. Our results connect to practical trajectory-based simulation strategies used in quantum transport \cite{BARKER2003, Oriols2007} and spin dynamics \cite{Dewdney1985,Brown1995,Das2019}, where understanding sampling efficiency matters.

The structure of this paper is organized as follows: In Sec. II, we review the guidance equations for spinless and spin-1/2 particles and clarify the role of equivariance. Section III clarifies the uncertainty in the trajectory framework and the meaning of finite-sample statistics. Section IV presents numerical results for the cases above, whose implications are analyzed and discussed in Sec. V. Finally, we conclude in Sec. VI.

\section{Bohmain mechanics and particle trajectories}
In this article, we restrict ourselves to the motion of a nonrelativistic particle.
\subsection{Single particle with spin-0}
The state of a spinless particle of mass $M$ subject to the external potential $V(\mathbf{x},t), \mathbf{x}=(x,y,z)$, is represented by the wavefunction, $\psi(\mathbf{x},t)$, whose time evolution is governed by the Schr\"{o}dinger equation,
\begin{equation}
    i\hbar\frac{\partial \psi(\mathbf{x},t)}{\partial t}=-\frac{\hbar^2}{2M}\nabla^2\psi(\mathbf{x},t)+V(\mathbf{x},t)\psi(\mathbf{x},t).
    \label{eq:schrodinger}
\end{equation}
In Bohmian mechanics, the wavefunction plays a dual role. First, by the Born rule, the square modulus $|\psi(\mathbf{x},t)|^2$ is the probability density function of detecting a particle at position $\mathbf{x}$ at time $t$ in one experiment. Meanwhile, the wavefunction also serves as a guidance field for the motion of a particle. This can be seen by writing the wavefunction in a polar form $\psi(\mathbf{x},t)=R(\mathbf{x},t)\exp(iS(\mathbf{x},t)/\hbar)$, substituting into the Schr\"{o}dinger equation and separating the imaginary and real parts, which gives two coupled equations:
\begin{align}
    \label{eq:quantumR2}
     &\frac{\partial \rho(\mathbf{x},t)}{\partial t}+\nabla\cdot\mathbf{j}=0,\\
    &\frac{\partial S(\mathbf{x},t)}{\partial t}+\frac{(\nabla S)^2}{2M}+V(\mathbf{x},t)+Q(\mathbf{x},t)=0,
    \label{eq:quantumHamiltonjacobi}
\end{align}
where $\rho\equiv R^2$,
\begin{equation}
    \mathbf{j}(\mathbf{x},t)=\rho(\mathbf{x},t)\frac{\nabla S(\mathbf{x},t)}{M}
\end{equation}
is the probability current defined through the velocity field
\begin{equation}
    \mathbf{v}(\mathbf{x},t)\equiv\frac{\nabla S(\mathbf{x},t)}{M},
    \label{eq:velo}
\end{equation} and
\begin{equation}
    Q(\mathbf{x},t)\equiv-\frac{\hbar^2}{2M}\frac{\nabla^2R}{R}
    \label{eq:quantumQ}
\end{equation}
is the quantum potential associated with the shape of the probability distribution \cite{Bohm1952}. The transport equation \eqref{eq:quantumR2} is a local conservation law for the probability flux constituted by an ensemble of trajectories resulting from individual, identical experiments. The quantum Hamilton-Jacobi equation  \eqref{eq:quantumHamiltonjacobi} incorporates the movement of the particle, whose equation of motion is
\begin{equation}
    M\frac{d^2\mathbf{x}}{dt^2}=-\nabla\left[ V(\mathbf{x},t)+ Q(\mathbf{x},t)\right].
    \label{eq:eomquantum}
\end{equation}
That is, the quantum particle moves under the action of a force that is not entirely derivable from the external potential $V(\mathbf{x},t)$; it also receives a contribution from the quantum potential $Q(\mathbf{x},t)$. The evolution of the wavefunction and hence the equation of motion \eqref{eq:eomquantum} is deterministic. To identify a unique trajectory, explicit initial (or instantaneous) values of position and velocity (momentum) must be specified. The quantum randomness in the context of Bohmian trajectories is manifested by the QEH \cite{Durr1992,Oriols2019}, which states that the position and velocity of a particular trajectory at an instant $t_0$ cannot be known with certainty. When the experiment is repeated $N(\rightarrow\infty)$ times, an ensemble of positions $\{\mathbf{x}^n[t_0]\}, n=1,...,N$ associated to the same $\psi(\mathbf{x},t_0)$ satisfy $\rho(\mathbf{x},t_0)=|\psi(\mathbf{x},t_0)|^2$. Alternative to Eq. \eqref{eq:velo}, the velocity of trajectory $n$ may be denoted as
\begin{equation}
    \mathbf{v}^n[t_0]=\frac{\mathbf{j}(\mathbf{x}^n[t_0],t_0)}{|\psi(\mathbf{x}^n[t_0],t_0)|^2},
    \label{eq:velo2}
\end{equation}
where
\begin{equation}
    \mathbf{j}(\mathbf{x},t)=i\frac{\hbar}{2M}\left[\psi(\mathbf{x},t)\nabla\psi^*(\mathbf{x},t)-\psi^*(\mathbf{x},t)\nabla\psi(\mathbf{x},t)\right]
    \label{eq:current}
\end{equation}
is the (ensemble) current density and $\psi^*$ is the complex conjugate of $\psi$. The QEH postulation guarantees that the quantum equilibrium ensemble of Bohmian trajectories produce the probability flux $\rho(\mathbf{x},t)=|\psi(\mathbf{x},t)|^2$ at any time, known as the equivariance \cite{Durr1992,Oriols2019}. Perhaps it is not surprising to view Eqs. (\ref{eq:quantumR2}-\ref{eq:eomquantum}) from the perspective of hydrodynamics that the streamlines of the probability flux are those particle trajectories, given that the initial position follows the probability distribution associated with the initial wavefunction.

\subsection{Single particle with spin-1/2}
Bohmian trajectories can also be denoted for particles carrying spins. For simplicity, we only consider the particle with spin-1/2. For a spin-$1/2$ charged particle (e.g. an electron), the complete (vectorial) wavefunction consists of two spinors $\Psi_{\uparrow,\downarrow}(\mathbf{x},t)$,
\begin{equation}
    \vec{\Psi}(\mathbf{x},t)=\left(\begin{matrix}
        \Psi_{\uparrow}(\mathbf{x},t)\\
        \Psi_{\downarrow}(\mathbf{x},t)
    \end{matrix}\right),
\end{equation}
governed by the Pauli equation,
\begin{equation}
    i\hbar\frac{\partial\vec{\Psi}}{\partial t}=\left[\frac{(-i\hbar\nabla-q\mathbf{A}/c)^2}{2M}-\hat{\boldsymbol{\mu}}\cdot\mathbf{B}+V\right]\vec{\Psi},
    \label{eq:SGHamil}
\end{equation}
with
\begin{equation}
    \mathbf{\mu}=\mu\boldsymbol{\sigma},\quad\mu\equiv\frac{q\hbar}{4Mc}g,
\end{equation}
where components of the vector $\boldsymbol{\sigma}=(\sigma^x,\sigma^y,\sigma^z)$ are Pauli matrices and $c$ is the speed of light. The value of the magnetic moment $\mu$ depends on the type of particle (e.g., electron, neutron, etc.). For an electron, it is in terms of mass $M$, charge $q$ and $g$-factor. The vector potential $\mathbf{A}$ and magnetic field $\mathbf{B}$ are related by
\begin{equation}
    \mathbf{B}=\nabla\times\mathbf{A}.
\end{equation}
%The interaction part in Eq. \eqref{eq:SGHamil} might be written as
%\begin{equation}
%    H_I=\epsilon(q)\sqrt{\alpha_q}\left[-\sqrt{\frac{\hbar}{c}}\mathbf{A}\cdot\mathbf{p}+\epsilon(q)\sqrt{\alpha_q}\frac{\hbar}{2Mc}\mathbf{A}^2-\frac{1}{4M}\sqrt{\frac{\hbar^3}{c}}g\mathbf{\sigma}\cdot\mathbf{B}\right],
%\end{equation}
%with $\alpha_a=q^2/(\hbar c)$ the dimensionless parameter and $\epsilon(q)=$sign$(q)$. Note that, for the electron $\alpha\approx 1/137$ is the fine-structure constant.
The continuity equation corresponding to the Pauli equation is
\begin{equation}
    \frac{\partial \rho(\mathbf{x},t)}{\partial t}+\nabla\cdot\mathbf{j}_{conv}(\mathbf{x},t)=0,
    \label{eq:currentspin}
\end{equation}
with
\begin{align}
\label{eq:densityspin}
    \rho(\mathbf{x},t)&=\vec{\Psi}^\dagger(\mathbf{x},t)\cdot\vec{\Psi}(\mathbf{x},t),\\
    \mathbf{j}_{conv}(\mathbf{x},t)&=i\frac{\hbar}{2M}\left(\vec{\Psi}(\mathbf{x},t)\cdot\nabla\vec{\Psi}^\dagger(\mathbf{x},t)-\vec{\Psi}^\dagger(\mathbf{x},t)\cdot\nabla\vec{\Psi}(\mathbf{x},t)\right)-\frac{q\mathbf{A}(\mathbf{x},t)}{mc}\rho(\mathbf{x},t),
\end{align}
the probability density and convective current, respectively \cite{holland1993}. The conjugate transpose of $\vec{\Psi}$ is denoted as $\vec{\Psi}^\dagger$ and similar to the spinless particle, the velocity can be denoted as 
\begin{equation}
    \mathbf{v}(\mathbf{x},t)=\frac{\mathbf{j}_{conv}(\mathbf{x},t)}{\rho(\mathbf{x},t)}.
    \label{eq:velospin}
\end{equation}
Spin components are expressed by $\mathbf{s}=\hbar\vec{\Psi}^\dagger\boldsymbol{\sigma}\vec{\Psi}/(2\rho)$. Note that because of the coupling of the magnetic field, spin can influence the dynamics of the particle. Another channel in which spin modifies trajectories is in the Pauli current, which will be discussed later.

\section{The uncertainty}
While Bohmian mechanics and standard quantum mechanics are probabilistic, there are some subtle differences between them. In the latter, the uncertainty (or variance) of an observable $\hat{A}$ in a state $\psi$ is defined by
\begin{align}
    (\Delta_\psi O)^2\equiv\int \psi^*\hat{O}^2\psi d^3x-\left(\int \psi^*\hat{O}\psi d^3x\right)^2.
\end{align}
It can be directly shown that there is a Heisenberg uncertainty principle for certain pairs of observables obeying the commutation relation \cite{Griffiths2018}. Alternatively, different observables could follow their own probability distributions. For instance, for position $\hat{x}_{j}$ and momentum $\hat{p}_k$ operators, it follows purely from the canonical commutation relation that $[\hat{x}_j,\hat{p}_k]=i\hbar\delta_{jk}$, where $\hbar$ is the Planck constant and $\delta_{jk}$ is the Kronecker delta. Given the state $\psi$, the uncertainty relation states that the product of uncertainties in position and momentum is bounded below by $\hbar/2$, 
\begin{equation}
    \Delta_\psi x_j\Delta_\psi p_k\geq\frac{\hbar}{2}\delta_{jk},\quad j,k=1,2,3.
    \label{eq:uncert}
\end{equation}
On the other hand, the wavefunction in momentum representation is the Fourier transform of the wavefunction in position representation, whose square amplitude is the probability distribution of momentum. The uncertainty relation becomes
\begin{equation}
    \Delta_\psi x_j\Delta_{\psi_p} p_k\geq\frac{\hbar}{2}\delta_{jk}.
    \label{eq:uncert2}
\end{equation}
where $\psi_p$ is the wavefunction in the momentum representation and
\begin{equation}
    \Delta_{\psi_p} p_k=\int|\psi_p|^2p_kd^3p.
\end{equation}

On the other hand, in the framework of Bohmian mechanics, since the wavefunction is conventionally expressed in position representation, the assignment of position enjoys the highest priority. Other quantities, such as velocity (momentum), spin, etc., are specified at a certain position. Besides, there is a gauge freedom in the definition of current (velocity): Adding a divergence-free vector field to the original current defines a current compatible with the same density of particles, c.f. Eq. \eqref{eq:quantumR2} \footnote{Adding a solenoidal component still keeps equivariance, and hence all operational uncertainty relations, such as Eq. \eqref{eq:uncert}, following from the Born rule, remain intact. However, the resulting equation of motion will have an additional potential depending on the velocity gauge. We shall continue to use the definition \eqref{eq:velo}.}. In all, the implication of the conventional uncertainty principle in the Bohmian context is not valid \footnote{What can change is the behavior of the unmeasured trajectories and “momentum”. Thus, what should be rephrased is the interpretation of uncertainty in the context of Bohmian trajectories. It has been suggested that the uncertainty relation in Bohmian mechanics is
\[\Delta_\psi x_j(\Delta_\psi p_{k})_{CL}\geq0,\]
with 
\[(\Delta p_{k})_{CL}=\int \rho(\partial S/\partial x_k)^2d^3x-\left(\int\rho(\partial S/\partial x_k)d^3x\right)^2\]
the single-valued classical ensemble of momentum \cite{holland1993}. }. To define Bohmian trajectories, the information of position should be specified a priori, at which values of other quantities are calculated, forming a unique trajectory.
%Note that the choice of position is directly due to the amplitude of the wavefunction, while the determination of the momentum is made through phase. It might be possible to introduce random variables to the phase of the wavefunction to maintain the conventional Heisenberg relation. However, it turns out that the relation is again violated in the subsequent evolution of the wavefunction. See the detailed discussions in appendix. 
In this regard, in the following sections, we will primarily investigate the statistics in positions and compare with the calculation from the wavefunction (probability distribution).

\section{Statistics in Bohmian trajectories}
First of all, as a direct implication of equivariance, the particle is static in eigenstates; the quantum potential balances the mechanical energy, leaving the state stationary. A similar conclusion can be drawn in the classical limit. When the probability distribution takes the form of a Dirac delta, the only possible way to place the particle is at the sharp of the distribution, at which the force induced by quantum potential vanishes and the particle is subject to the guidance of the external potential, forming the classical trajectory \cite{Cui2023,Cui2025}.

In the following examples, we will track trajectories exclusive to noneigenstates. 

\subsection{The harmonic oscillator}
\begin{figure}[!tp]
\includegraphics[width=0.8\textwidth]{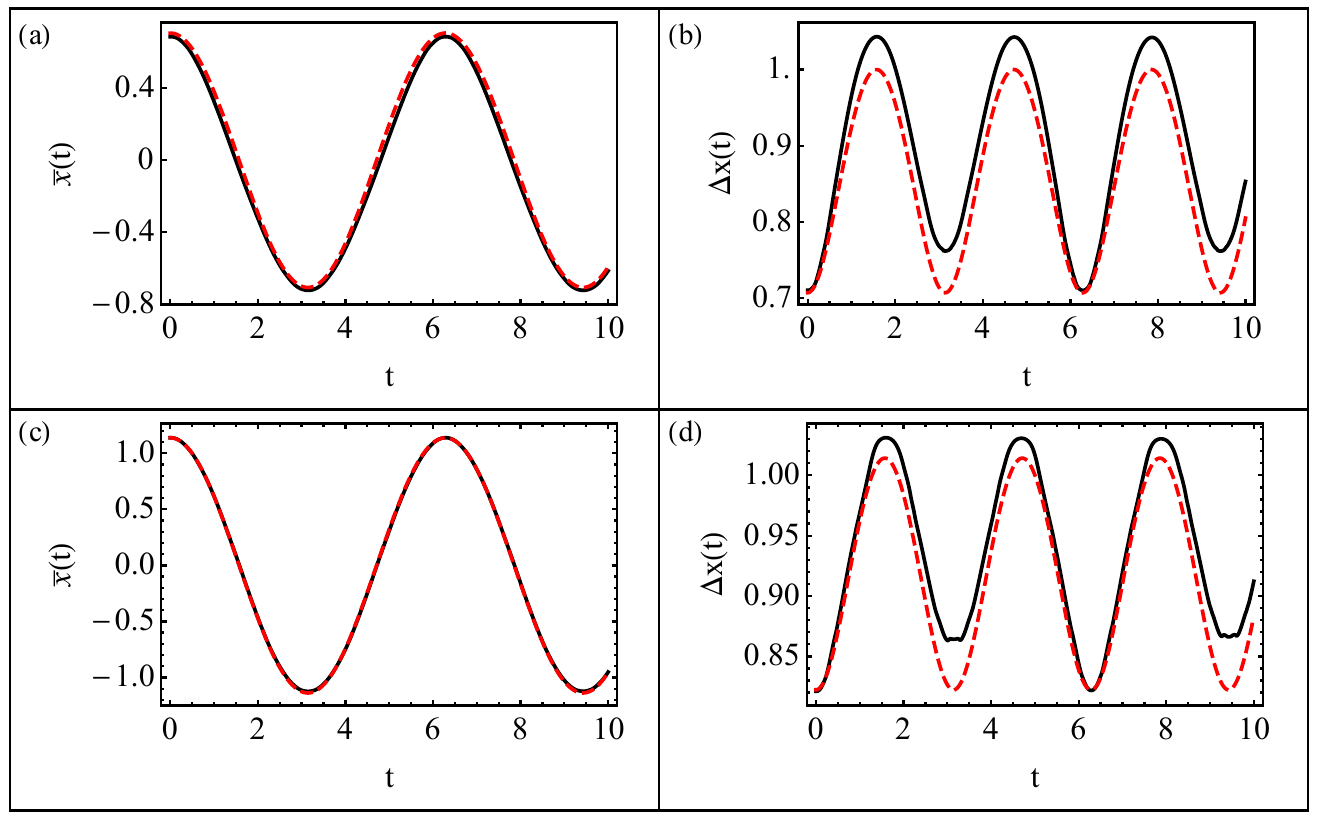}
\caption{The motion of a quantum harmonic oscillator with mass $M$ and frequency $\omega$, released from a superposition of a few eigenstates. The initial state in panels (a) and (b) is given by Eq. \eqref{eq:harsup1}; in panels (c) and (d), by Eq. \eqref{eq:harsup2}.  Panels (a) and (c) indicate the mean position, while (b) and (d) present the standard deviation of position, as functions of time. In all panels, the black solid lines indicate the results of statistical analysis based on 400 trajectories, whereas the red dashed lines correspond to calculations employing the probabilistic character of the wavefunction. Time and displacement are measured in units of $\omega^{-1}$ and $(\hbar/M\omega)^{-1/2}$, respectively.}
\label{fig:1}
\end{figure}

We consider a single quantum harmonic oscillator of mass $M$ and frequency $\omega$ in 1D. The initial wavefunction is taken as a linear superposition of two or three eigenstates,
\begin{equation}
    \psi_1(x,0)=\frac{e^{-\frac{\alpha^2x^2}{2}}}{\sqrt{2/\alpha}}\left(\frac{H_0(\alpha x)}{\sqrt{\pi^{1/2}2^00!}}+\frac{H_1(\alpha x)}{\sqrt{\pi^{1/2}2^11!}}\right)
    \label{eq:harsup1}
\end{equation}
or
\begin{equation}
    \psi_2(x,0)=\frac{e^{-\frac{\alpha^2x^2}{2}}}{\sqrt{3/\alpha}}\left(\frac{H_0(\alpha x)}{\sqrt{\pi^{1/2}2^00!}}+\frac{H_1(\alpha x)}{\sqrt{\pi^{1/2}2^11!}}+\frac{H_2(\alpha x)}{\sqrt{\pi^{1/2}2^22!}}\right),
    \label{eq:harsup2}
\end{equation}
where $H_n(x)$ are Hermite polynomials of degree $n$ and $\alpha=\sqrt{M\omega/\hbar}$. As the quantum potential changes in time, particles are subjected to move under the effect of the quantum potential and the external quadratic potential. In Fig. \ref{fig:1}, we show plots of the mean and standard deviation of position from a collection of particle trajectories launched from different initial states superposed by a few eigenstates, indicating a good agreement with the 1st and 2nd moments of the probabilistic wavefunction.

\begin{figure}[!tp]
\includegraphics[width=0.8\textwidth]{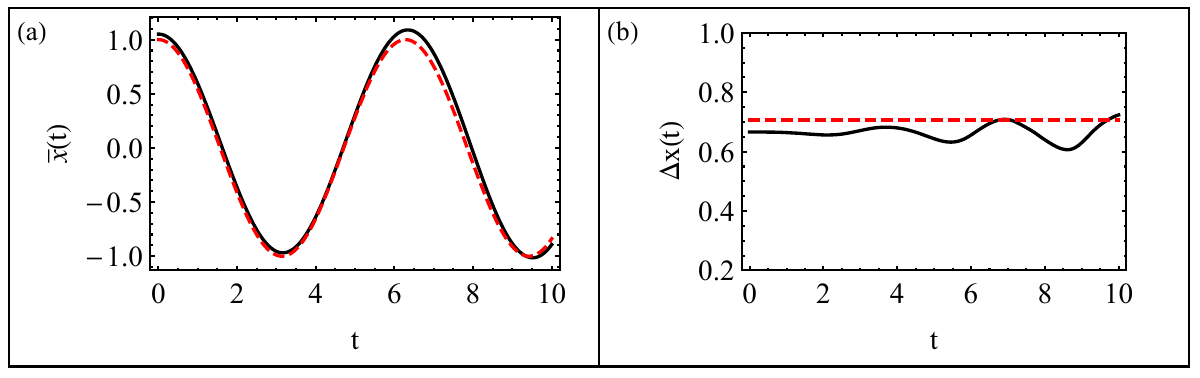}
\caption{The motion of a quantum harmonic oscillator with mass $M$ and frequency $\omega$, released from a wavepacket centered at $x_c=1$ with the spread $\gamma^2=0.5$ and zero phase, c.f. Eq. \eqref{eq:harmpt}. Panels (a) and (b) show the time dependence of the mean and the standard deviation of position, respectively. In all panels, the black solid lines indicate the results of statistical analysis based on 400 trajectories, whereas the red dashed lines correspond to calculations employing the probabilistic character of the wavefunction. Time and displacement are measured in units of $\omega^{-1}$ and $(\hbar/M\omega)^{-1/2}$, respectively.}
\label{fig:2}
\end{figure}

We also take a wavepacket for the initial state, a typical shape in the quantum-to-classical transition \cite{Cui2023}. The time evolution of the wavefunction takes the form
\begin{subequations}
\begin{align}
    \psi(x, t;x_c,p_0, \gamma^2)&=A\exp\left[-a(t)(x-q(t))^2+\frac{i}{\hbar}p(t)(x-q(t))\right],\\
    a(t)&=\frac{M\omega}{\hbar}\frac{\hbar\cos(\omega t)+i2\gamma^2M\omega\sin(\omega t)}{i2\hbar\sin(\omega t)+4\gamma^2M\omega\cos(\omega t)},\\
    q(t)&=x_c\cos(\omega t)+\frac{p_0}{M\omega}\sin(\omega t),\\
    p(t)&=p_0\cos(\omega t)-M\omega x_c\sin(\omega t),
\end{align}
\label{eq:harmpt}
\end{subequations}
where $A$ is a normalization constant, $\gamma$ is the initial spread, and we have neglected the purely time-dependent phase that has no physical interest. The position and momentum of the peak in the packet at $t=0$ are $x_c$ and $p_0$, respectively. The wavepacket (or probability distribution) sustains the Gaussian shape, and 
\begin{align}
    \text{Re}[a(t)]&=\frac{1}{4\gamma^2[\cos^2(\omega t)+\tan^2\phi\sin^2(\omega t)]},
\end{align}
with
\begin{align}
    \tan\phi&=\frac{\hbar}{2\gamma^2M\omega}.
\end{align}
From Eq. \eqref{eq:eomquantum}, the quantum force is
\begin{align}
    F_Q&\equiv-\frac{\partial Q(x,t)}{\partial x}\\
    &=\frac{\hbar^2(x-q(t))}{4\gamma^4M[\cos^2(\omega t)+\tan^2\phi\sin^2(\omega t)]^2},
\end{align}
which oscillates in phase as the spread of $R(x,t)$ when the particle does not follow the classical trajectory $q(t)$. In particular, if the oscillatory frequency is $\omega=\hbar/(2\gamma^2M)$, the spread $a$ is constant in time. That is, the oscillator is in a coherent state. There, the quantum force $F_Q$ maintains a constant shift $\Delta x$ from the classical trajectory $q(t)$ when the initial position of the particle is detected by $\Delta x=x-x_c$ away from the peak $x=x_c$. Suppose a particle is located at $x_0$ at $t=0$, its Bohmian trajectory is
\begin{equation}
    X[t]=x_0+x_c\left[\cos(\omega t)-1\right]+\frac{p_0}{M\omega}\sin(\omega t).
    \label{eq:harmox}
\end{equation}
Since the initial position $x_0$ follows the normal distribution $\sim N(x_c,\gamma^2)$, by virtue of equivariance, statistics of $X[t]$ will constitute the same distribution as the coherent state with the peak shifted to $q(t)$, which is shown in Fig. \ref{fig:2}. The same investigations are also obtained for wavepackets with other initial widths. See Fig. \ref{fig:S1} in Appendix A for details.

\subsection{Some ``exceptions"}
Results presented in the last section are not surprising because the equivariance ensures that the probabilistic distributions are indeed formed by particle flows, yielding the same predictions. Rigorously, however, a series of infinite identical experiments should be implemented to capture the complete probability distribution. In practice, it is only possible to count a finite number of Bohmian trajectories, which might cause noticeable differences.

\begin{figure}[!tp]
\includegraphics[width=0.8\textwidth]{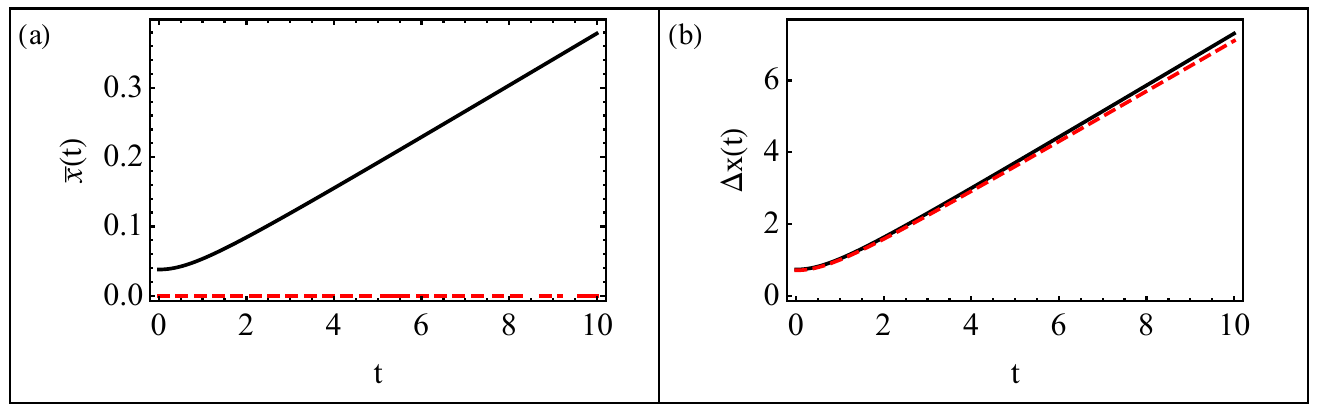}
\caption{The motion of a free particle with a unit mass $M$, released from a wavepacket centered at $x_c=0$ with spread $\gamma^2=0.5$ and zero phase. Panels (a) and (b) show the time dependence of the mean and standard deviation of position, respectively. In all panels, the black solid lines indicate the results of statistical analysis based on 400 trajectories, whereas the red dashed lines correspond to calculations employing the probabilistic character of the wavefunction. In panel (b), black and red lines overlap. The Planck constant is taken to be $\hbar=1$ .}
\label{fig:3}
\end{figure}

We first pick a free particle of mass $M$ released from a wavepacket centered at $x=0$ with zero phase. In a finite number of experiments, albeit close, the initial mean position does not necessarily coincide with the origin, and a time-dependent (non-zero) quantum force will drive it to move, causing the expansion of the wavepacket. As the spread of the wavepacket grows in time, the (mean) trajectory will move further away from $x=0$ at a later time, as is seen in Fig. \ref{fig:3}(a). However, a good agreement is still attained if we compare the standard deviation, which is shwon in Fig. \ref{fig:3}(b). Similar results are also obtained for the superposition of wavepackets. See Fig. \ref{fig:S2} in Appendix A. In Fig. \ref{fig:S3}, we also show the case for the harmonic oscillator with the same superpositional wavepackets.

\begin{figure}[!tp]
\includegraphics[width=0.8\textwidth]{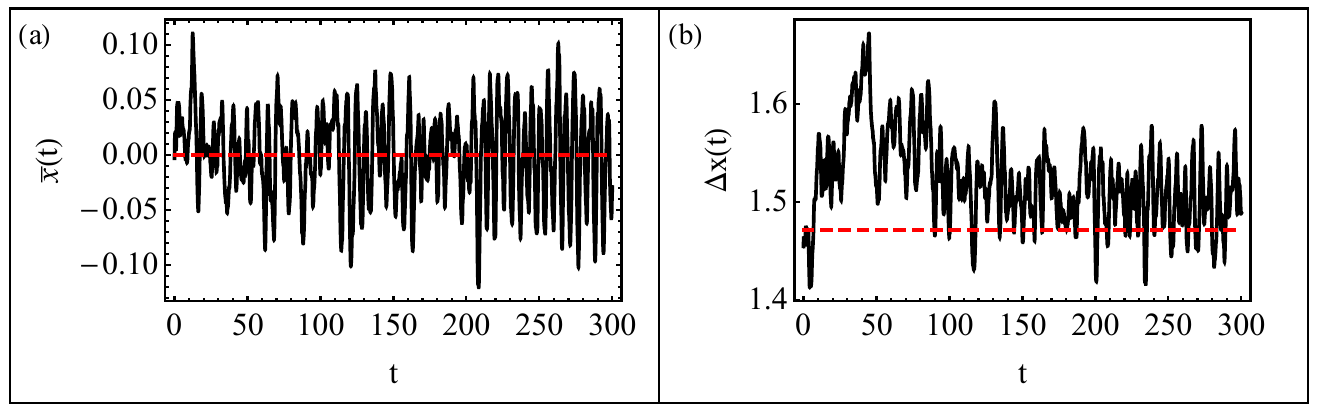}
\caption{The motion of a 3D quantum harmonic oscillator with mass $M$ and oscillatory frequency $\omega$, released from an initial state in Eq. \eqref{eq:harm3D}. Panels (a) and (b) show the time dependence of the mean and standard deviation of position, respectively. In all panels, the black solid lines indicate the results of statistical analysis based on 250 trajectories, whereas the red dashed lines correspond to calculations employing the probabilistic character of the wavefunction. Time and displacement are measured in units of $\omega^{-1}$ and $(\hbar/M\omega)^{-1/2}$, respectively.}
\label{fig:4}
\end{figure}

Another situation where statistics of a collection of finite trajectories might lead to an apparent difference is when Bohmian trajectories exhibit chaos \cite{Cushing2000,Efthymiopoulos2006}. We consider a quantum harmonic oscillator in 3D, whose Hamiltonian is
\begin{equation}
    \hat{H}=-\frac{\hbar^2\nabla^2}{2M}+\frac{M\omega^2}{2}(\hat{x}^2+\hat{y}^2+\hat{z}^2).
\end{equation}
The initial state is taken to be a linear combination of a few degenerate eigenstates,
\begin{equation}
    \psi_{3D}(x,y,z)=\frac{1}{\sqrt{3}}\left(\phi_{1,1,1}(x,y,z)+e^{i\pi/3}\phi_{3,0,0}(x,y,z)+e^{i\pi/7}\phi_{1,2,0}(x,y,z)\right),
    \label{eq:harm3D}
\end{equation}
where
\begin{equation}
    \phi_{n_x,n_y,n_z}(x,y,z)=\frac{e^{-(\alpha^2x^2+\alpha^2 y^2+\alpha^2z^2)/2}H_{n_x}(\alpha x)H_{n_y}(\alpha y)H_{n_z}(\alpha z)}{\sqrt{(\pi/\alpha)^{3/2}2^{n_x+n_y+n_z}n_x!n_y!n_z!}},
\end{equation}
have eigenenergy 
\begin{equation}
    E_{n_x,n_y,n_z}^{3d}=\hbar\omega\left(n_x+n_y+n_z+\frac{3}{2}\right).
\end{equation}
The Lyapunov exponent is 0.06; the particle's motion depends sensitively on the initial position  \cite{Cesa2016}. Discernible departures in the average and standard deviation of position from predictions of the (probabilistic) wavefunction can be observed in Fig. \ref{fig:4} (and also Fig. \ref{fig:S4} in Appendix A for movements along other directions). Interestingly, there is a nodal plane at $x=0$, to which velocity becomes tangential and Bohmian trajectories do not cross, while probabilistic calculations yield that the expectation value of the position should lie at $x=0$ \footnote{There are other nodal lines ending in the plane $x=0$. See \cite{Cesa2016} for details.}. We thus remark that the average trajectory calculated at each instant is not necessarily the trajectory of the mean position at the initial time, since the latter may not even exist.

\subsection{Dynamics with spins}
\begin{figure}[!tp]
\includegraphics[width=0.7\textwidth]{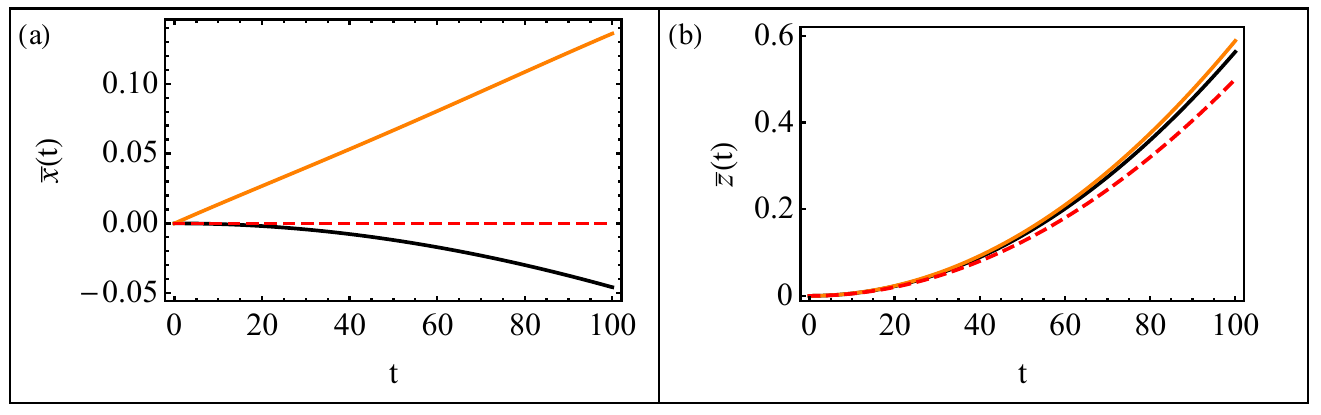}
\caption{The motion of a spin-1/2 neutral particle with a unit mass subjected to external time-dependent magnetic field $\mathbf{B}=(0,-by,bz)$, released from a wavepacket centered at the origin with spread $\gamma^2=100$ in all directions, whose momentum of the $y$-component of the wavepacket is $0.08$ while other components are zero. The spin part is aligned with the -$z$-axis. Panels (a) and (b) show the time dependence of the mean position along the $x$- and $z$-axes, respectively, where the starting points of all curves are shifted to the origin. In all panels, the black and orange solid lines are the statistical analysis based on 400 trajectories of convective and Pauli currents, respectively. Calculations employing the probabilistic character of the wavefunction are represented by red dashed lines. The magnetic field $b$ is chosen to be 0.1 and the magnetic moment is $\mu=-0.001$. Atomic units are used in the calculation unless otherwise specified.}
\label{fig:5}
\end{figure}

We attempt to construct Bohmian trajectories of particles carrying spins by looking at the dynamics under the magnetic field $\mathbf{B}(t)=(0,-by,bz)$, where $b>0$ is a constant. Such a configuration of a magnetic field is typical in Stern-Gerlach experiments \cite{Batelaan1997,Manoukian2003}. For convenience, we assume that the particle is neutral so that the interaction is between spin and magnetic field. The initial state is separable into an orbital part $\psi(\mathbf{x},0)$ taken to be a 3D Gaussian wavepacket,
%with
%\begin{equation}
%    \gamma(t)=\gamma\left(1+\frac{\hbar^2t^2}{4M^2\gamma^4}\right)^{1/2},
%\end{equation}
plus the spin part $\vec{\chi}(0)=(A \quad B)^T$ with $A^2+B^2=1$. If the strength of the magnetic field is weak, $b\ll1$, up to the order $\sim\mathcal{O}(b)$, the time evolution of wavefunction can be approximated as
\begin{align}
    \vec{\Psi}(\mathbf{x},t)=\psi(\mathbf{x},t)\left(\begin{matrix}
        \chi_{\uparrow}\\
        \chi_{\downarrow}
    \end{matrix}\right)
    \label{eq:wavefSG}
\end{align}
with
\begin{subequations}
\begin{align}
\label{eq:free3DGauss}
    \psi(\mathbf{x},t)&\propto\exp\left[-\frac{(\mathbf{x}-\mathbf{p}_0t/M)^2}{4\gamma(t)^2}+\frac{i\mathbf{p}_0\cdot\mathbf{x}}{\hbar}\right],\\
    \chi_{\uparrow}&=A-\frac{\mu bB}{\hbar}\left[yt-\frac{t^2}{2M}\left(p_y+\frac{i\hbar y'}{2\gamma^2(t)}\right)\right]+\frac{i\mu bA}{\hbar}\left[zt-\frac{t^2}{2M}\left(p_z+\frac{i\hbar z'}{2\gamma^2(t)}\right)\right],\\
    \chi_{\downarrow}&=B+\frac{\mu bA}{\hbar}\left[yt-\frac{t^2}{2M}\left(p_y+\frac{i\hbar y'}{2\gamma^2(t)}\right)\right]-\frac{i\mu bB}{\hbar}\left[zt-\frac{t^2}{2M}\left(p_z+\frac{i\hbar z'}{2\gamma^2(t)}\right)\right],
\end{align}
where 
\begin{equation}
    \gamma(t)=\gamma\left(1+\frac{\hbar^2t^2}{4M^2\gamma^4}\right)^{1/2}
\end{equation}
\end{subequations}
is the spread, $y'=y-p_yt/M,z'=z-p_zt/M$ and $\mathbf{p}_0=(p_x,p_y,p_z)$. Compare to a free particle, from Eq. \eqref{eq:wavefSG}, it can be calculated that, up to $\mathcal{O}(b)$, only the mean value of $z$ will receive a correction quadratic in $t$, $\mu bt^2(A^2-B^2)/(2M)$, whereas means along other directions and variances of positions are the same. In Fig. \ref{fig:5}, we present statistics of motion for particles whose initial spin is aligned with the $-z$-axis, $\vec{\chi}(0)=(0 \quad 1)^T$, and the initial wavefunction carries momentum $p_y>0, p_x=p_z=0$. A good agreement between the statistics of Bohmian trajectories and the Born rule is investigated in particular for the convective current at a short time. Motions along other directions are depicted in Fig. \ref{fig:S5} in Appendix A.

%Since the Lorentz force causes an obvious deviation of the particle from its initial path, resulting in a blurring in the beam splitting, it has been suggested to set a non-uniform magnetic field so that the intensity distribution on the screen reflects non-trivial correlations between dynamical variables \cite{Batelaan1997,Manoukian2003}. For example, the vector $A$ can be chosen as $\mathbf{A}=(bz,\beta xz,\beta xy)$, giving the according magnetic field $\mathbf{B}=(0,b-\beta y, \beta z)$. The conventional $z$-component is the conventional quantization axis of spin, whereas the $y$-component of the magnetic field, $b-\beta x_2$, is aligned with the direction of propagation. The uniform part $(0,b,0)$ of the magnetic field is longitudinal. It can be appropriately tuned to effectively reduce the correlation between $x_1$ and $x_3$ variables on screen. The remaining part $(0,-\beta x_2,\beta x_3)$ is almost longitudinal since the macroscopic distance $|x_2|$ is much larger than $|x_3|$ that provides a splitting of the beam. 

It has been suggested to associate the current vector with an additional spin (Pauli) current \cite{Colijn2002,Holland2003},
\begin{equation}
    \mathbf{j}_{tot}=\mathbf{j}_{conv}+\frac{\hbar}{2M}\nabla\times(\vec{\Psi}^\dagger\boldsymbol{\sigma}\vec{\Psi}),
    \label{eq:totcurrvelo}
\end{equation}
which gives a different definition of velocity, while maintaining the same continuity equation, Eq. \eqref{eq:currentspin}. As a consequence, the statistics on the trajectories might be different, as is shown in Fig. \ref{fig:5}. Further, in Fig. \ref{fig:S6} in Appendix B, we show results of another model to reveal differences in trajectories from convective current and Pauli current.

%However, we might not form trajectories with Eq. \eqref{eq:totcurrvelo} because the Pauli equation essentially consists of four consistent equations: the Hamilton-Jacobi-type equation indicating the dynamics of the particle, the continuity equation and two equations for two spin components (orientations), enforcing the definition of velocity as Eq. \eqref{eq:velospin} \cite{holland1993}.

\section{Discussion}
Bohmian dynamics are deterministic once initial conditions are specified; statistical predictions arise from ensembles of initial positions distributed by $|\psi|^2$ and transported by the guidance flow (equivariance). The results reported in this article should therefore be read as an analysis of finite-sample effects: how estimates formed from a practical number $N$ of trajectories approximate the Born rule moments, and when deviations can be visible for realistic $N$.

In regular guidance regimes—Gaussian wavepacket and low-mode superpositions of eigenstates of the harmonic oscillator—sample means and variances converge rapidly to the quantum predictions. Figures \ref{fig:1}–\ref{fig:2} (and Fig. \ref{fig:S1} in Appendix A) illustrate that, for modest $N$ (hundreds of trajectories), the first and second moments of position closely track those computed from the wavefunction. Small biases in the sample mean can occur when the initial draw of positions is slightly off-centered, which would diminish with increasing $N$ and do not affect variances appreciably. However, sometimes the slight deviation in mean position due to finite sample statistics could be more pronounced at long times. This is seen in Fig. \ref{fig:3} (and Fig. \ref{fig:S2} in appendix), where the movement of the free particle is only driven by the quantum potential, while the variance indicates no obvious bias (also see Fig. A3 in appendix).

Moreover, for degenerate superpositions in 3D harmonic oscillators with nodal lines/surfaces and chaotic guidance, the flow partitions configuration space into nodal cells, and trajectories exhibit sensitive dependence on initial conditions. In such cases (Fig. \ref{fig:4} and Fig. \ref{fig:S4} in appendix), finite-sample estimates of the mean and standard deviation can display noticeable deviations from the Born rule expectation. This behavior reflects two intertwined factors: (i) poor mixing across nodal cells (trajectories cannot cross nodes), which can bias sampling if initial positions underrepresent some cells; and (ii) chaotic stretching within cells, which amplifies small imbalances in the initial ensemble. It is important to note that these are sampling issues tied to the flow structure, not a failure of equivariance: As $N\rightarrow\infty$ with appropriate sampling across cells, ensemble statistics coincide with $|\psi|^2$.

For spin-1/2 dynamics, both the convective and the Pauli (total) currents generate continuity equations that preserve $|\psi|^2$, but they define different velocity fields. Our comparisons (Fig. \ref{fig:5} and Figs. A5–A6 in appendix) show that, for finite $N$, trajectory statistics obtained from these two currents can differ—most clearly in certain mean and variance estimates—while ensemble equivalence in the quantum equilibrium limit remains intact. Practically, the choice of current can influence finite-sample behavior and should be stated explicitly in trajectory-based simulations of spin transport \cite{Dewdney1985,Brown1995,Das2019} or Stern–Gerlach-type setups \cite{Krekels2024,Gondran2014}.

We emphasize that any visible discrepancy in our figures is a finite-sample effect compounded by flow geometry. In the quantum equilibrium ensemble, Bohmian mechanics reproduces orthodox statistics exactly. The practical question is how large $N$ must be—and how to sample efficiently—to reach the desired accuracy for a given dynamical regime.

\section{Conclusion}
In summary, we studied how finite ensembles of Bohmian trajectories approximate standard quantum statistics across regimes ranging from regular to chaotic guidance flows. In some regular cases (e.g. Gaussian wavepackets, low-mode superpositions of eigenstates), sample means and/or variances rapidly match Born-rule predictions for modest number of trajectories. In contrast, for selected degenerate superpositions of the 3D harmonic oscillator with nodal barriers and chaotic guidance, finite-sample means can exhibit noticeable deviations. This may offer a new perspective on computing particle trajectories in surface hopping \cite{Curchod2013}. For spin-1/2 particles, both convective and Pauli currents guarantee equivariance; however, 
their distinct velocity fields can produce different finite-sample trajectory statistics, even though the equilibrium ensemble remains unchanged.

Our findings do not challenge the equivalence between Bohmian mechanics and orthodox quantum mechanics in quantum equilibrium. Instead, they highlight, that (i) finite-sample deviations are expected and can be amplified by flow features such as nodal partitioning and chaos, and (ii) careful sampling strategies and diagnostics (e.g., stratification across nodal cells, convergence monitoring) are advisable for trajectory-based simulations, especially in spin-transport contexts.

We also note that both Bohmian mechanics and standard quantum mechanics assume the validity of the Born rule, which has been questioned as a possible approximation to some underlying (deterministic) theory from which quantum mechanics emerges \cite{Engel1992,Landsman2022}. Performing statistics in Bohmian trajectories paves the way to test the validity of quantum equilibrium. Finally, in this article, we limit our discussion to the first and second moments of position in single-particle systems. Extending the analysis to many-body systems (with entanglement and conditional wave functions), to other observables modeled via explicit measurement interactions, and to rigorous convergence-rate analyses are natural next steps.\color{black}

\section*{Acknowledgements}
Discussions with Zhouhao Wang are appreciated. B.C. acknowledges the financial support of the National Natural Science Foundation of China (No. 12404232), start-up funding from the Chinese University of Hong Kong, Shenzhen (No. UDF01003468) and the Shenzhen city “Pengcheng Peacock” Talent Program.

\section*{Author contributions}
B. C. designed the project and analyzed the data. B. C and Y. C. performed calculations and wrote the paper.

\section*{Data availability statements}
No data was used for the research described in the article.

\begin{appendix}

\section{More figures}
\renewcommand{\thefigure}{A\arabic{figure}}

\setcounter{figure}{0}
In this section, we present additional figures supplementary to the results discussed in the main text.

\begin{figure}[!tp]
\includegraphics[width=0.8\textwidth]{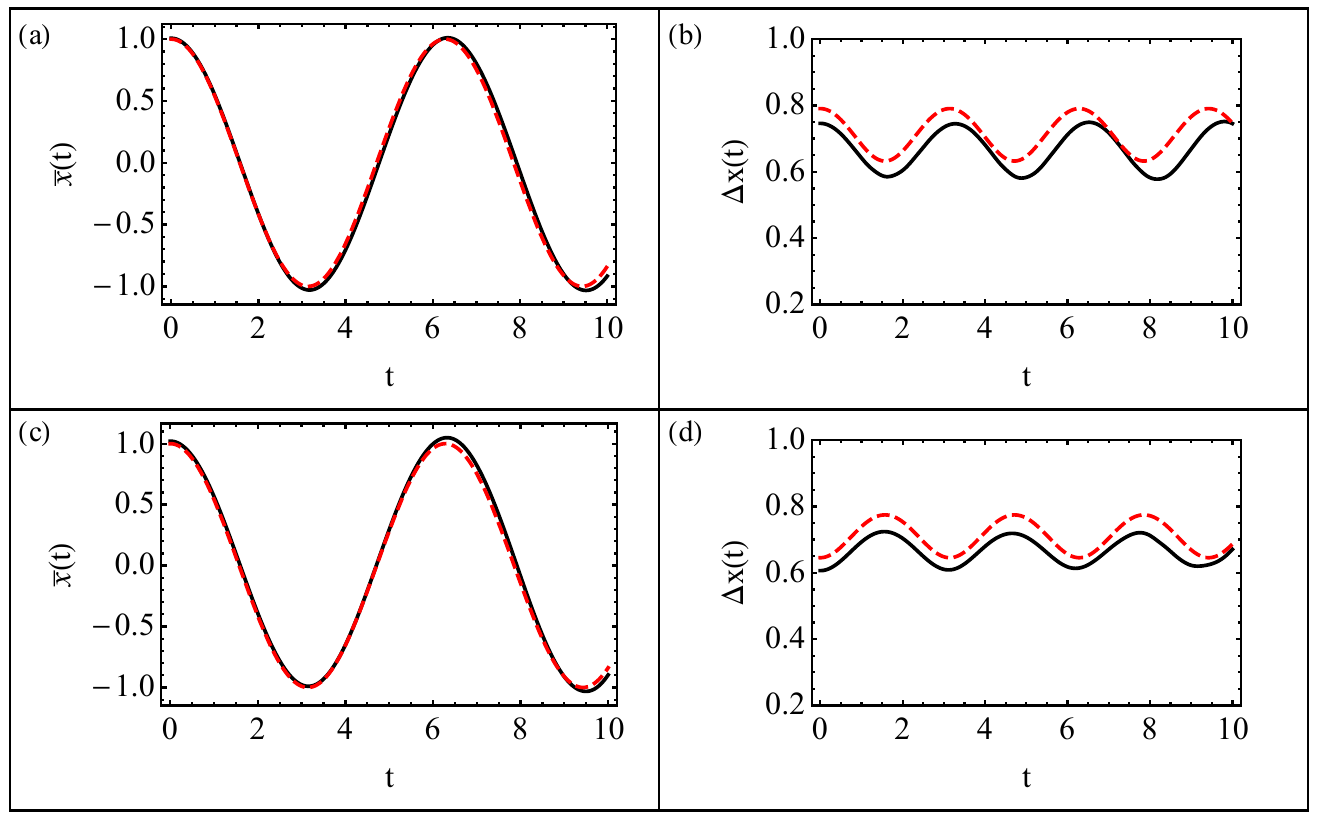}
\caption{The motion of a quantum harmonic oscillator with mass $M$ and frequency $\omega$, released from a wavepacket centered at $x_c=1$ with zero phase, c.f. Eq. \eqref{eq:harmpt}. The initial spread $\gamma$ in panels (a-b) is $\gamma^2=1/1.6$, while in panels (c-d), it is $1/2.4$. Panels (a) and (c) show the time dependence of the mean position; panels (b) and (d) display the standard deviation. In all panels, the black solid lines indicate the results of statistical analysis based on 400 trajectories, whereas the red dashed lines correspond to calculations employing the probabilistic character of the wavefunction. Time and displacement are measured in units of $\omega^{-1}$ and $(\hbar/M\omega)^{-1/2}$, respectively.}
\label{fig:S1}
\end{figure}

In Fig. \ref{fig:S1}, we show the mean and standard deviation of Bohmian trajectories of a quantum harmonic oscillator from wavepackets with various variances.

\begin{figure}[!tp]
\includegraphics[width=0.8\textwidth]{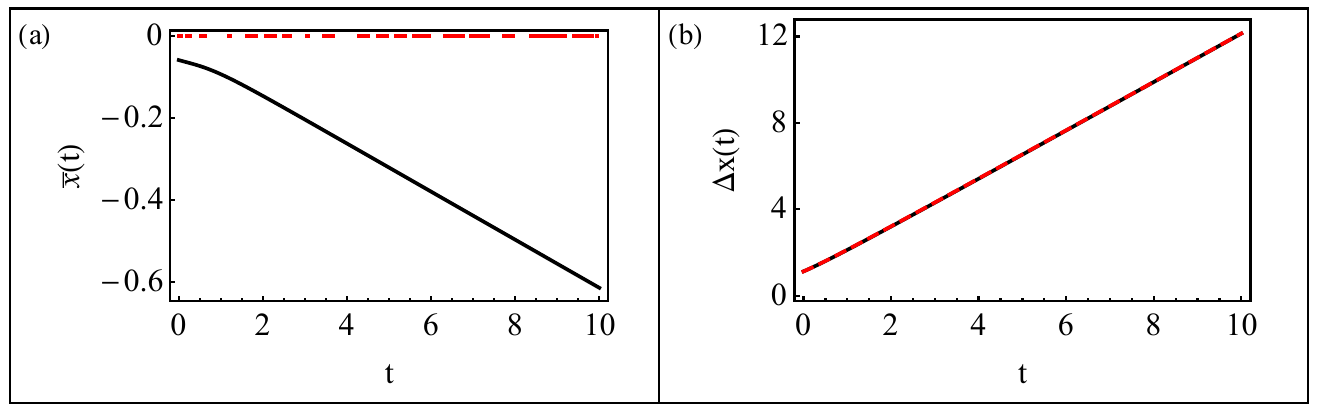}
\caption{The motion of a free particle with a unit mass $M$, released from a superposition of two wavepackets, c.f. Eq. \eqref{eq:harmsup}, with $x_c=1$ and $\gamma^2=0.5$. Panels (a) and (b) show the time dependence of the mean and standard deviation of position, respectively. In both panels, the black solid lines indicate the results of statistical analysis based on 400 trajectories, whereas the red dashed lines correspond to calculations employing the probabilistic character of the wavefunction. In panel (b), black and red lines overlap. Planck constant is taken to be $\hbar=1$.}
\label{fig:S2}
\end{figure}

\begin{figure}[!tp]
\includegraphics[width=0.8\textwidth]{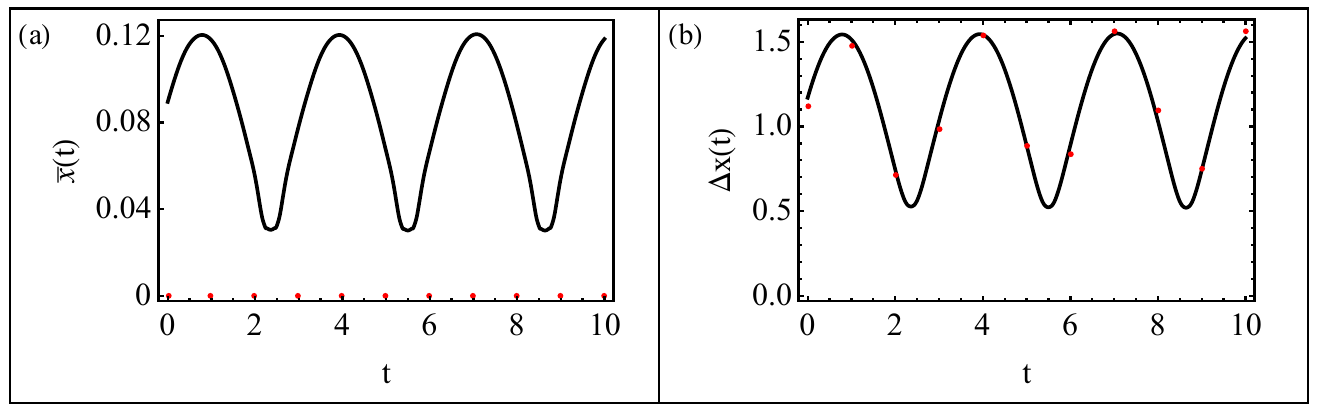}
\caption{The motion of a quantum harmonic oscillator with mass $M$ and frequency $\omega$, released from a superposition of two wavepackets, c.f. Eq. \eqref{eq:harmsup}, with $x_c=1$ and $\gamma^2=0.5$. Panels (a) and (b) show the time dependence of the mean position and standard deviation. In both panels, the black solid lines indicate the results of statistical analysis based on 400 trajectories, whereas the red dots correspond to calculations employing the probabilistic character of the wavefunction. Time and displacement are measured in units of $\omega^{-1}$ and $(\hbar/M\omega)^{-1/2}$, respectively.}
\label{fig:S3}
\end{figure}

Figure \ref{fig:S2} displays the statistics in Bohmian trajectories of a free particle with a superposition of two wavepackets initially centered at $x=\pm x_c$ and moving in opposite directions, i.e.,
\begin{align}
    \psi(x,t)=\mathcal{N}\left(\exp\left[-\frac{(x-x_c)^2}{4\gamma^2}+\frac{ip_0x}{\hbar}\right]+\exp\left[-\frac{(x+x_c)^2}{4\gamma^2}-\frac{ip_0x}{\hbar}\right]\right),
    \label{eq:harmsup}
\end{align}
where $\mathcal{N}$ is a normalization constant, while results for a particle subject to a quadratic potential from the same initial state are depicted in Fig. \ref{fig:S3}.

\begin{figure}[!tp]
\includegraphics[width=0.8\textwidth]{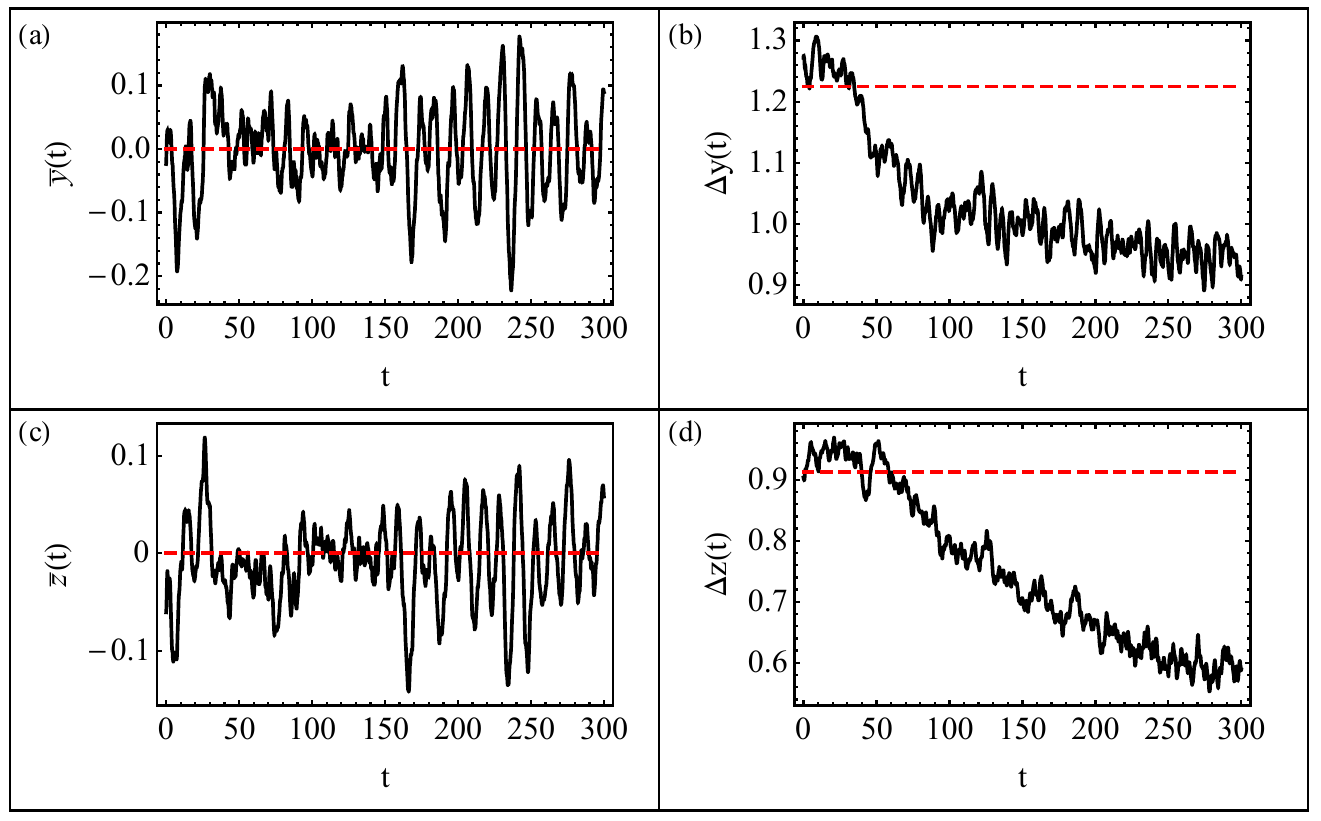}
\caption{The motion of a quantum harmonic oscillator in 3D with mass $M$ and oscillatory frequency $\omega$, released from an initial state Eq. \eqref{eq:harm3D} in the main text. Panels (a) and (c) show the time dependence of the mean position; panels (b) and (d) display the standard deviation. In all panels, the black solid lines indicate the results of statistical analysis based on 250 trajectories, whereas the red dashed lines correspond to calculations employing the probabilistic character of the wavefunction. Time and displacement are measured in units of $\omega^{-1}$ and $(\hbar/M\omega)^{-1/2}$, respectively.}
\label{fig:S4}
\end{figure}

\begin{figure}[!tp]
\includegraphics[width=0.7\textwidth]{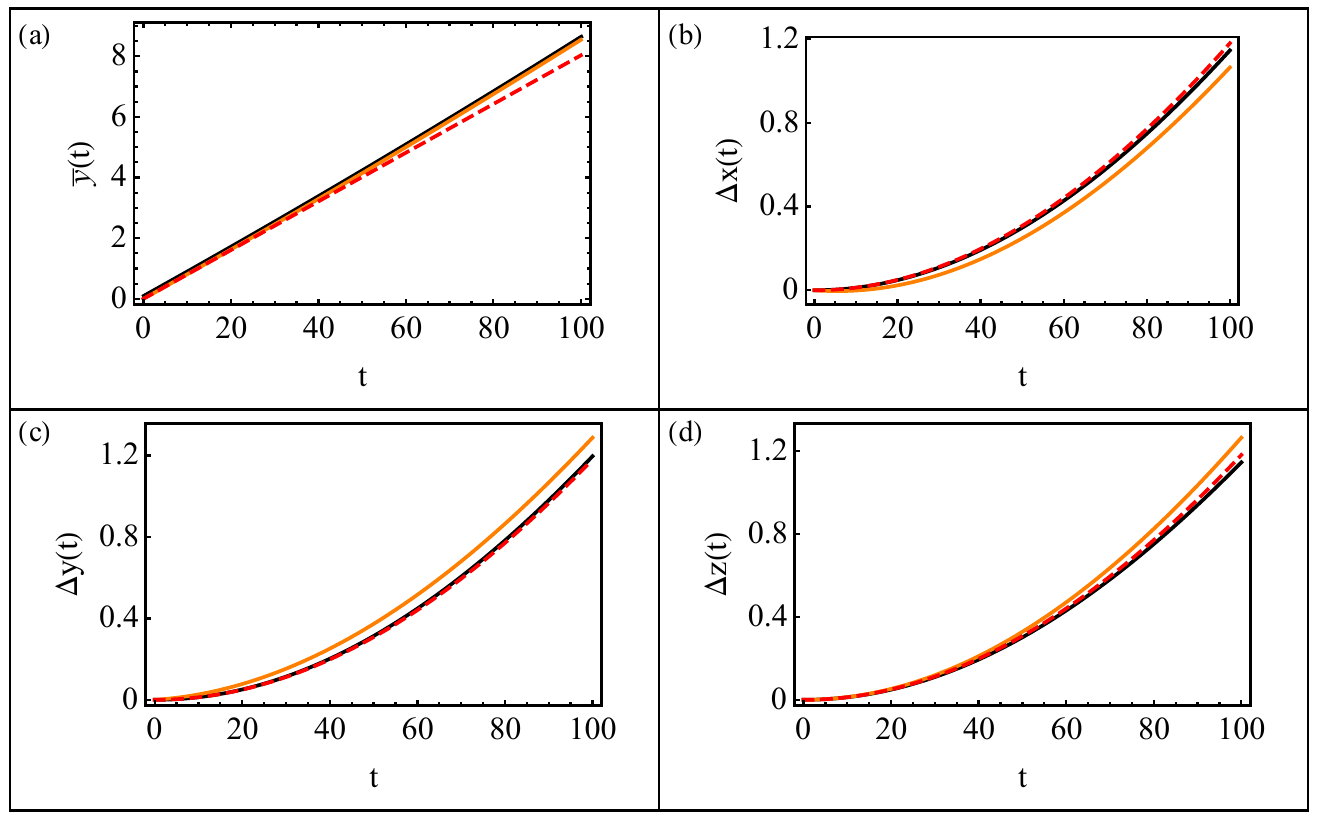}
\caption{The motion of a spin-1/2 neutral particle with a unit mass subjected to external time-dependent magnetic field $\mathbf{B}=(0,-by,bz)$, released from a wavepacket centered at the origin with spread $\gamma^2=100$ in all directions. The momentum of the $y$-component of the wavepacket is taken $0.08$ while other components are zero. The spin part is aligned with the -$z$-axis. Panels (a) show the time dependence of the mean position along the $y$-axis, while panels (b-d) depict the standard deviation along all directions. The starting points of all curves are shifted to the origin. In all panels, the black and orange solid lines are the statistical analysis based on 400 trajectories of convective and Pauli currents, respectively. Calculations employing the probabilistic definition of the wavefunction are represented by red dashed lines. The magnetic field $b$ is chosen to be 0.1 and the magnetic moment is $\mu=-0.001$. Atomic units are implemented in the calculation unless otherwise specified.}
\label{fig:S5}
\end{figure}

Chaotic trajectories of a 3D harmonic oscillator along other directions, which is discussed in Sec. IVB in the main text, are displayed in Fig. \ref{fig:S4}.

In Fig. \ref{fig:S5}, we depict means and standard deviations of motion supplementary to the spin-1/2 particle discussed in the main text.

\begin{figure}[!tp]
\includegraphics[width=0.7\textwidth]{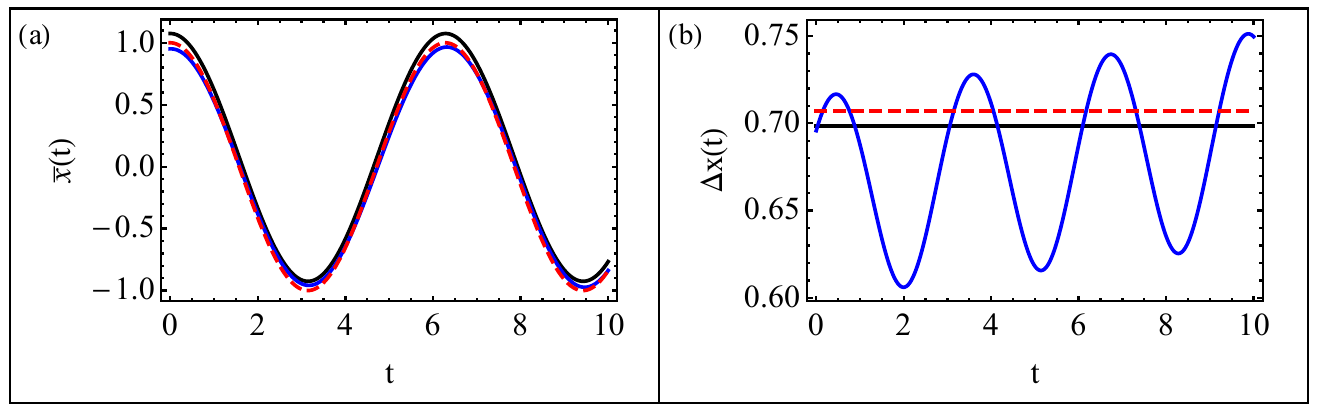}
\caption{The motion of a spin-1/2 neutral quantum harmonic oscillator with a unit mass $M$ and frequency $\omega=1$ subjected to external time-dependent magnetic field $\mathbf{B}(t)=(B\cos(\omega t),B\sin(\omega t),B_0)$, released from a wavepacket centered at the origin with spread $\gamma^2=0.5$ in all three directions. The momentum of the y-component of the wavepacket is taken $0.08$ while other components are zero. The spin part is aligned with the +$z$-axis. Panels (a) and (b) show the time dependence of the mean and standard deviation of position along the $x$-axis, respectively. In all panels, the black and blue solid lines are the statistical analysis based on 150 trajectories of convective and Pauli currents, respectively. Calculations employing the probabilistic definition of the wavefunction are represented by red dashed lines. The static magnetic field $B_0$ is 1 Tesla, while the strength of the oscillatory field is $B=0.001$ Tesla with frequency $\omega=7.06\times10^{-10}$. The magnetic moment is $\mu=-0.001$. Atomic units are implemented in the calculation unless otherwise specified.}
\label{fig:S6}
\end{figure}

\section{Spin dynamics in a time-dependent magnetic field}
In this section, we consider the motion of a spin-1/2 neutral harmonic oscillator under a superposition of an oscillatory and uniform magnetic field $\mathbf{B}(t)=(B_1\cos(\omega t),B_1\sin(\omega t),B_0)$, where $B$ and $B_0$ are constants. The setup of the magnetic field is common in Ramsey experiments of separated fields \cite{Manoukian2006}. Note that the interaction term is purely time-dependent. The total wavefunction is separable into orbital and spin parts. The orbital part agrees with the Gaussian wavepacket of a free particle, c.f. Eq. \eqref{eq:free3DGauss} in the main text. The unitary evolution operator for the spin part is \cite{Manoukian2006}
\begin{align}
    U(0,t)=\left(\begin{matrix}
        U_{11}(t) &U_{21}(t)\\
        U_{21}(t) &U_{22}(t)
    \end{matrix}\right),
\end{align}
where
\begin{align}
    U_{11}(t)&=\left[\cos(at)+i\left(\frac{\omega-\omega_0}{2a}\right)\sin(at)\right]e^{-i\omega t/2},\\
    U_{12}(t)&=-i\frac{|\mu|B_1}{\hbar a}\sin(at)e^{-i\omega t/2},\\
    U_{21}(t)&=-i\frac{|\mu|B_1}{\hbar a}\sin(at)e^{i\omega t/2},\\
    U_{22}(t)&=\left[\cos(at)-i\left(\frac{\omega-\omega_0}{2a}\right)\sin(at)\right]e^{i\omega t/2},\\
    a&=\left[\left(\frac{\omega-\omega_0}{2}\right)^2+\frac{\mu^2B_1^2}{\hbar^2}\right]^{1/2},\\
    \omega_0&=2|\mu|B_0/\hbar.
\end{align}
In Fig. \ref{fig:S6}, we compare the results of trajectories calculated from convective and Pauli currents with the initial spin aligned with +$z$-axis. Surprisingly, the mean paths agree well with probabilistic calculations, while the standard deviation of positions in Pauli currents experiences oscillations.

\end{appendix}

\bibliography{reference}

\end{document}